\documentstyle[11pt,paspconf,psfig]{article}

\def\point{   \parindent=10pt \par
              \hangindent\parindent  \textindent  }
\font\ninerm=cmr9    \font\sixrm=cmr5
\font\nineit=cmti9  
\font\ninesl=cmsl9
\font\ninei=cmmi9    \font\sixi=cmmi5
\skewchar\ninei='177
\font\ninesy=cmsy9  \font\sixsy=cmsy5
\skewchar\ninesy='60
\font\ninebf=cmbx9  \font\sixbf=cmbx5
\font\nineex=cmex10 scaled 833

\font\ninett=cmtt9
\def\adjustlinespace{\baselineskip=\baselineskip}
\def\ninepoint{\textfont0=\ninerm \scriptfont0=\sixrm 
                \def\rm{\fam0\ninerm}\relax
                \textfont1=\ninei \scriptfont1=\sixi 
                \def\mit{\fam1}\def\oldstyle{\fam1\ninei}\relax
                \textfont2=\ninesy \scriptfont2=\sixsy 
                \def\cal{\fam2}\relax
                \textfont3=\nineex \scriptfont3=\nineex 
                \def\it{\fam\itfam\nineit}\relax
                \textfont\itfam=\nineit
                \def\sl{\fam\slfam\ninesl}\relax
                \textfont\slfam=\ninesl
                \def\bf{\fam\bffam\ninebf}\relax
                \textfont\bffam=\ninebf \scriptfont\bffam=\sixbf
                \def\tt{\fam\ttfam\ninett}\relax
                \textfont\ttfam=\ninett
                \setbox\strutbox=\hbox{\vrule
                     hnine7pt depth2pt width0pt}\baselineskip=9pt
                \adjustlinespace
                \rm}
\font\caps=cmcsc10			   
\def\etal{{\it et~al.\ }}
\def\aa #1 #2 {A\&A, #1, #2}
\def\aas #1 #2 {A\&AS, #1, #2}
\def\acm #1 #2 {ACM-Trans Math Software, #1, #2}
\def\ada #1 #2 {Ann Astrophys, #1, #2}
\def\agabstr #1 #2 {Astr Ges Abstr Ser, #1, #2}
\def\aj #1 #2 {AJ, #1, #2}
\def\anach #1 #2 {Astr Nachr, #1, #2}
\def\apj #1 #2 {ApJ, #1, #2}
\def\apjl #1 #2 {ApJL, #1, #2}
\def\apjs #1 #2 {ApJS, #1, #2}
\def\araa #1 #2 {ARAA, #1, #2}
\def\apss #1 #2 {ApSpaceS, #1, #2}
\def\celmech #1 #2 {Cel Mech, #1, #2}
\def\esom #1 #2 {ESO Messenger, #1, #2}
\def\fundcp #1 #2 {FunCosP, #1, #2}
\def\jcp #1 #2 {J Comp Phys, #1, #2}
\def\jfm #1 #2 {J Fluid Mech, #1, #2}
\def\jmp #1 #2 {J Math Phys, #1, #2}
\def\ma #1 #2 {Mitt Astr Ges, #1, #2}
\def\mn #1 #2 {MNRAS, #1, #2}
\def\nat #1 #2 {Nat, #1, #2}
\def\obs #1 #2 {Observatory, #1, #2}
\def\pasj #1 #2 {PASJ, #1, #2}
\def\pasp #1 #2 {PASP, #1, #2}
\def\phyr #1 #2 {PhysRep, #1, #2}
\def\physd #1 #2 {Physica D, #1, #2}
\def\rpp #1 #2 {RepProgPhys, #1, #2}
\def\ssr #1 #2 {Sp Sci Rev, #1, #2}
\def\iau127#1{in de Zeeuw P.T. ed, Structure and Dynamics of 
     Elliptical Galaxies, IAU Symp.~No.~127. Reidel, Dordrecht, p.~#1}

\def\in#1#2#3#4#5#6{in: #1%
\if#2-%
\else%
, #2%
\fi%
\if#3-%
\else%
, ed.\ #3%
\fi%
\if#5-%
 {\if#4-%
 \else,%
   (#4)%
 \fi}%
\else%
 {\if#4-%
, (#5)%
\else%
, (#5:#4)%
\fi}%
\fi%
\if#6-%
.%
\else%
, #6.%
\fi%
}

\def\spose#1{\hbox to 0pt{#1\hss}}
\def\lta{\mathrel{\spose{\lower 3pt\hbox{$\mathchar"218$}}
     \raise 2.0pt\hbox{$\mathchar"13C$}}}
\def\gta{\mathrel{\spose{\lower 3pt\hbox{$\mathchar"218$}}
     \raise 2.0pt\hbox{$\mathchar"13E$}}}
\def\equal{\! = \!}

\def\=#1{\overline{#1}}

\def\df{{\caps df}}
\def\vp{{\caps vp}}

\def\hfifty{h_{50}}

\def\kms{{\rm\,km\,s^{-1}}}

\def\kpc{{\rm\,kpc}}

\def\msun{{\rm\,M_\odot}}

\def\rms{{\caps rms}}

\def\ni{\noindent}

\begin{document}

\title{Dynamical Mass Determination for Elliptical Galaxies}

\author{Ortwin Gerhard, Gunther Jeske}
\affil{Astronomisches Institut, Universit\"at Basel, Venusstrasse 7, CH-4102
       Binningen, Switzerland}
\author{R.~P.~Saglia, Ralf Bender}
\affil{Universit\"ats--Sternwarte, Scheinerstr. 1, D-81679 M\"unchen, Germany}

\begin{abstract}
The mass and anisotropy of an elliptical galaxy can be simultaneously
determined from velocity dispersion and line profile shape
measurements. We describe the principles, techniques, and limitations
of this approach, and the results obtained sofar. We briefly discuss
how best to combine these stellar-dynamical results with X-ray
measurements and gravitational lensing analyses.
\end{abstract}

\null\vspace*{-0.8cm}

\section{Introduction}

To place elliptical galaxies in the context of galaxy formation
theories, it is vital to understand their mass distributions and
dynamical structure. Yet despite nearly two decades of work we are
only beginning to answer important questions such as:
\point{$\triangleright$}
What are the density distribution, the flattening, the extent of
elliptical galaxy halos?  Are these similar as for the halos of spiral
galaxies, for the same stellar mass?
\point{$\triangleright$}
To what extent has the luminous matter in ellipticals dissipatively
segregated in the dark matter potential? Is there a ``conspiracy''
between luminous and dark matter to produce a flat circular orbit
rotation curve, like in spiral galaxies?
\point{$\triangleright$}
How anisotropic, radially or tangentially, are the velocity
ellipsoids?
\point{$\triangleright$}
How do the derived halo properties and dynamical structure compare
with those obtained from numerical simulations of the merger of two
spiral galaxies? Can the results be
used to constrain the r\^ole of the baryons in hierarchical merging?

\ni
Progress in this field has been beset by well--known difficulties:
There is no single ideal tracer like a spiral galaxy's rotation curve,
and the interpretation of velocity dispersion data is ambigous,
suffering from a fundamental degeneracy between anisotropy and mass.
However, recent advances on several fronts give reason for optimism:
\point{$\triangleright$}
From absorption {\sl line profile} data it is now possible to
determine {\sl both} the mass distribution {\sl and} anisotropy out
to, at present, $\lta 2-3$ half--light radii $R_e$. How this can be
done, and what the results are to date, is the subject of this
article. The method works, in principle, for all nearby ellipticals.
\point{$\triangleright$}
The improved spatial and spectral resolution of ROSAT and ASCA has
allowed accurate mass determinations to large radii for several
massive ellipticals (Mushotzky \etal 1994, Kim \& Fabbiano 1995).
Resolution limits the analysis to $R\gta R_e$. The number of
X--ray bright ellipticals analysed in this way is still relatively
small.
\point{$\triangleright$}
Galaxy--galaxy (strong) lensing gives accurate surface mass
measurements at a specific radius defined by the lensed images (e.g.,
Kochanek \& Keeton 1997). Galaxy--galaxy weak lensing (Brainerd,
Blandford \& Smail 1996) will in future allow a statistical
determination of outer surface mass profiles for galaxies at typical
redshift of a few tenths.

\ni
Both X--ray and gravitational lensing mass measurements will be
described elsewhere in these proceedings.  In this article we explain
the stellar dynamical principles of how line profile data can break
the traditional degeneracy (Section 2).  We show that the analysis of
real data does indeed work (Section 3), and then describe the results
obtained sofar (Section 4). It is important to note that combining
results from several approaches will be very useful; thus we briefly
comment on how the new dynamical determinations fit in with the
results from other techniques (Section 5).

\section{Mass and anisotropy from line profile data: Principles}

Illustrative line--of--sight velocity profiles (\vp s) for isotropic,
radially, and tangentially anisotropic distribution functions in fixed
potentials are given, e.g, in Dejonghe (1987), Merritt (1987) and
Gerhard (1991) for spherical systems, and in Dehnen \& Gerhard (1993,
1994), Evans (1993), and Qian \etal\ (1995) for axisymmetric
systems. The signature of radial orbits is a broad central profile,
and a peaked profile with extended wings beyond the half--mass radius,
where the profile is dominated by orbits near tangent point, the point
along the line--of--sight closest to the galaxy center.  Tangential
orbits tend to generate narrow profiles in the center and broad,
sometimes double--peaked profiles at large radii.  In axisymmetric
potentials, rotation introduces a noticeable asymmetry with respect to
the \vp\ center, and the minor and major axis anisotropies can be
decoupled.

\begin{figure*}[t]
\vspace*{0cm}
\centering
\hfil\psfig{figure=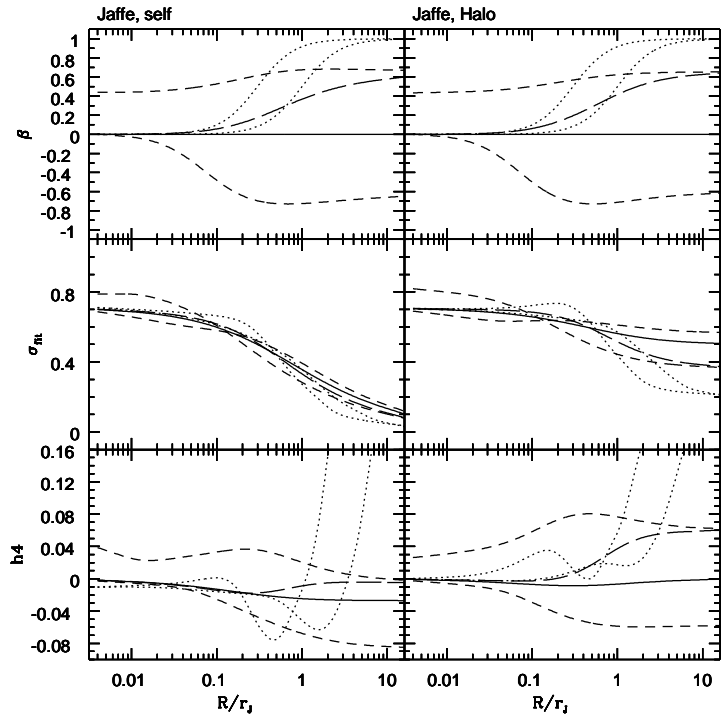,width=10cm}\hfil
 \vspace*{-0.6cm}
 \caption[models]
    {\ninepoint Anisotropy parameter $\beta$, projected velocity
	dispersion $\sigma_{\rm fit}$, and \vp-parameter $h_4$ for
	representative Jaffe models in self-consistent (left) and halo
	potential (right). $\sigma_{\rm fit}$ and $h_4$ are obtained
	by simultaneously fitting a Gaussian and a 4$^{\rm th}$--order
	Gauss-Hermite function to the \vp.  The models shown are the
	isotropic model (solid), radially and tangentially anisotropic
	models constructed with the method of Gerhard (1991) (dashed
	lines), and two radially anisotropic Osipkov (1979)-Merritt
	(1985) models (dotted lines). Note that while $\beta$ is a
	function of three-dimensional radius $r$ and $\sigma_{\rm
	fit}$, $h_4$ are observed quantities depending on projected
	radius $R$, there is a close correspondence between features
	in these profiles. }
 \label{models}
\end{figure*}

While the anisotropy of an elliptical galaxy is thus clearly
constrained by \vp\ shape measurements, it can not be directly
inferred because the \vp s also depend on the gravitational potential
(Gerhard 1993).  An updated illustration for spherical Jaffe models of
various anisotropy structure is given in Fig.~\ref{models}.  The \vp\
shapes are conveniently represented by Gauss--Hermite moments (Gerhard
1993, van der Marel \& Franx 1993).  In the spherical case the \vp\ is
symmetric, and is parametrized in Fig.~\ref{models} by the {\sl
fitted} profile width and lowest order even coefficient $h_4$ (van der
Marel \& Franx 1993). Fig.~\ref{models} shows a general trend: As the
mass of the model at large $r$ is increased at constant anisotropy,
both the projected dispersion and $h_4$ increase. Increasing $\beta$
at constant potential, on the other hand, lowers $\sigma$ and
increases $h_4$.  This suggests that by modelling $\sigma$ and $h_4$
both mass $M(r)$ and anisotropy $\beta(r)$ can in principle be found.

Locally, this trend may be reversed, as shown by the two
Osipkov-Merritt-models in Fig.~\ref{models}; in this case a large
number of high-energy radial orbits all turning around near radius
$r_a$ lead to flat-topped \vp s in a small radial range near
$r_a$. Although the properties of these models are extreme and they
have not been very useful sofar for modelling observed \vp s, this
nonetheless suggests that a global non-parametric analysis of the data
is required.

\section{Analysis of observed line--profile data}

In recent studies, analyzing simulated and real line-profile data has
proceeded in two distinct steps (Merritt 1993, Rix \etal 1997, Gerhard
\etal 1997): First, the gravitational potential $\Phi$ is held fixed,
and in this current potential a best--estimate \df\ is determined from
the data. Then, this process is repeated for different potentials, and
confidence intervals are found for the parameters entering the
functional form of $\Phi$.

In the first step, the \df\ is written as a sum of elements (which can
be values in grid cells, orbits, or basis functions). These elements
are projected into observational space, and their weights are adjusted
to fit the data. The equation relating the \df\ to the {\sl projected}
\df\ (over line-of-sight velocities and sky positions) is linear, so
this is relatively straightforward. Also, at least in the spherical
case we know that the solution is unique for ideal data (Dejonghe \&
Merritt 1992). The same is likely to be true for edge-on axisymmetric
systems, as long as only one orbit family is involved.  Because real
data provide only a sparse and noisy subset of the projected \df, only
gross features of the phase-space \df\ can be extracted, and
significant smoothing is necessary. This is done by the concept of
regularisation (see Merritt 1993). For each set of data an optimal
regularisation parameter $\lambda$ must be determined.

Fig.~\ref{simu} shows the application of such a technique to Monte
Carlo data, which is obtained from a model \df\ in a known potential,
but using the sampling and radial extent of real data for the E0
galaxy NGC 6703 (Gerhard \etal 1997). 

\newdimen\halfwidth \halfwidth=.48\hsize

\begin{figure}[hb]
\vspace*{-.1cm}
\hbox to \hsize{\psfig{file=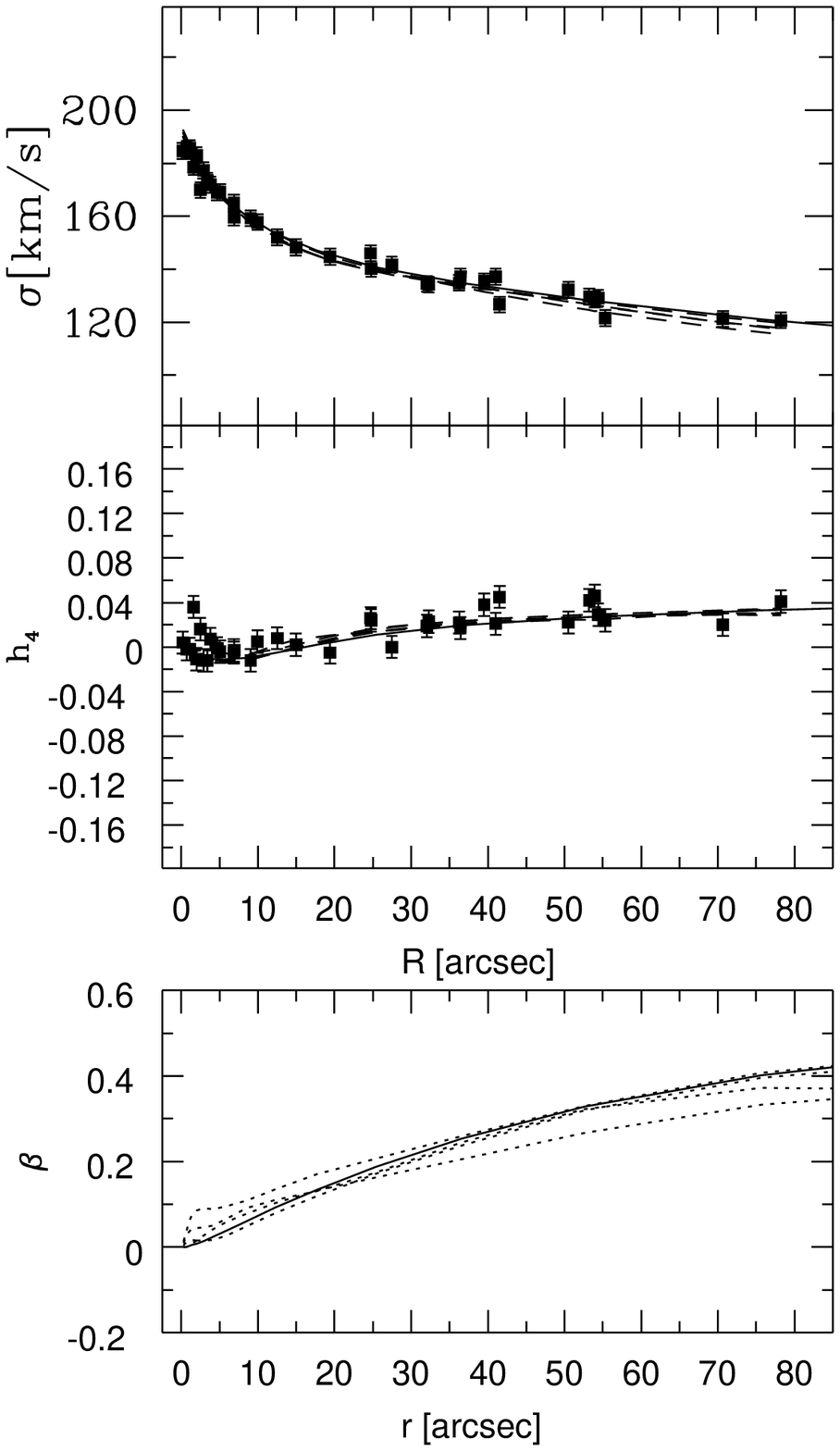,width=\halfwidth} \hfill
		\psfig{file=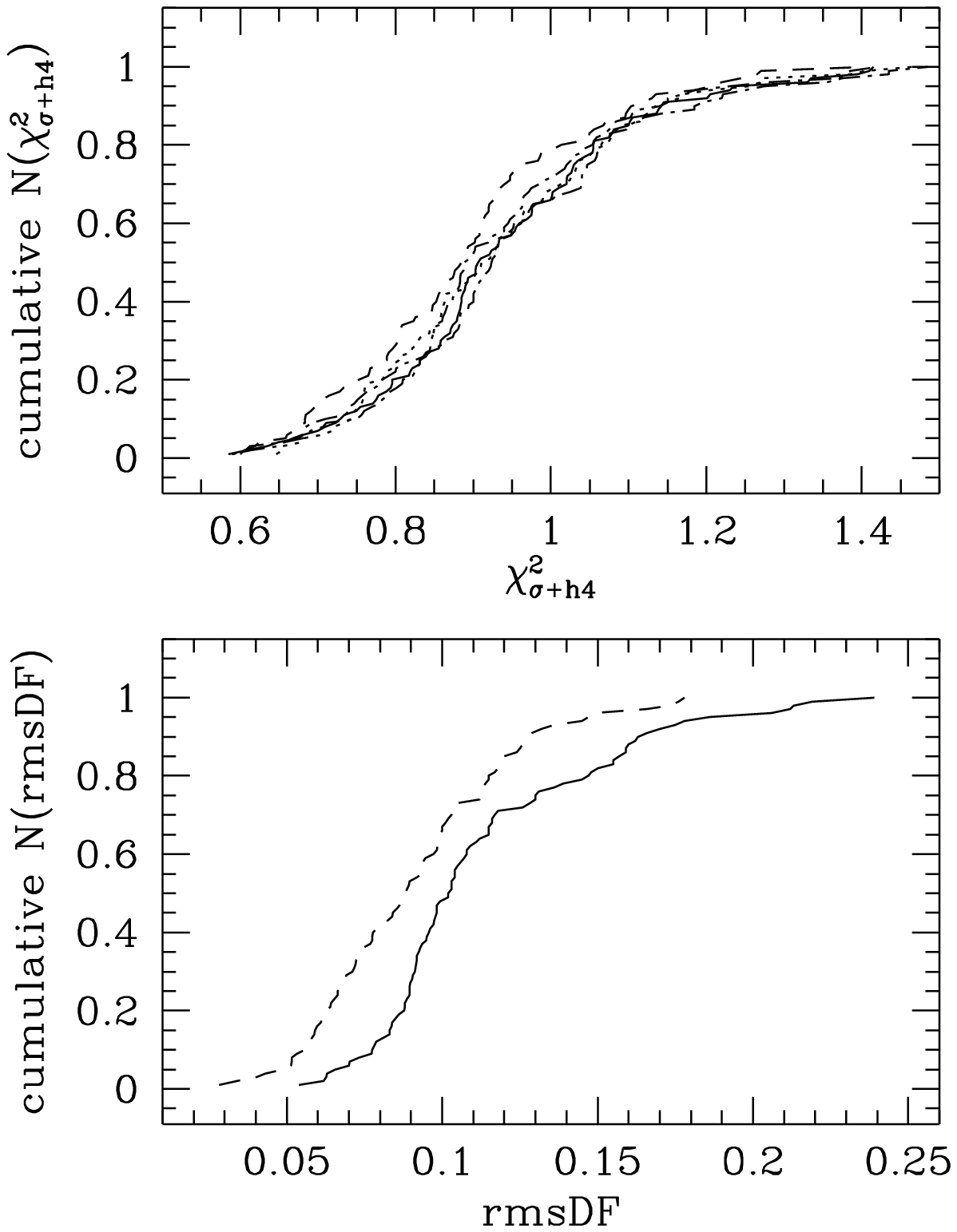,width=\halfwidth} }
 \vspace*{-0.1cm}
 \caption[kinlambdam]
    {(Left): \ninepoint Model analysis of pseudo-data generated from a
radially anisotropic model \df. The sampling and radial extent are the
same as for the real data of NGC 6703 (Fig.~\ref{n6703}), but assuming
error bars $\Delta\sigma\equal 3\kms$ and $\Delta h_4\equal 0.01$.
The full curves show the true $\sigma$, $h_4$, and $\beta$ profiles of
the underlying \df. The dashed lines show the projected $\sigma$ and
$h_4$ profiles derived from the pseudo-data by our method for optimal
$\lambda$.  The dotted curves show the uncertainties that remain in
the deprojected quantity $\beta$ even with such small error bars.}
 \label{simu}
 \vspace*{-0.2cm}
 \caption[distrchisq]
   {(Right): \ninepoint Top: The cumulative distribution of the
normalized $\chi^2_{\sigma+h4}$, for dynamical models recovered from
100 random Gaussian data sets derived from the same \df, with
observational errors as in Fig.~\ref{simu} (dashed line) and
Fig.~\ref{n6703} (full).  Bottom: Cumulative distribution of residuals
between true and recovered distribution function, evaluated on a grid
extending to three times the radius of the last data point, for the
same 100 datasets. Also shown in the top panel are the cumulative
$\chi^2_{\sigma+h4}$--distributions for data drawn from radially
anisotropic models in the two potentials corresponding to two of the
extreme solid lines in Fig.~\ref{allvc}, with the same error bars as
in Fig.~\ref{n6703} (dot--dashed lines), and for a self--consistent
model with more complicated anisotropy structure (dotted line).}
 \label{distrchisq}
\end{figure}

\newpage\ni
The assumed sizes of the error
bars in this example are $3\kms$ for the velocity dispersion $\sigma$,
and $0.02$ for the \vp\ shape parameter $h_4$. The elements used in
this particular technique are various isotropic and tangentially
anisotropic basis functions constructed along the lines described by
Gerhard (1991).  Fig.~\ref{simu} shows that the true \df\ can be
recovered well from such data, and Fig.~\ref{distrchisq} shows
quantitatively the cumulative distribution of $\chi^2$ per data point
for random Monte Carlo realizations. In this example, the algorithm
will match the data with $\chi^2_{\sigma+h4}<1.28$ with $95\%$
confidence, and recover an underlying smooth \df\ to within an \rms\
of $12\%$.

The second step, that of constraining the potential, is harder because
$\Phi$ enters non-linearly into the equation for the projected
\df. Moreover, experiments suggest that the potential is in practice
determined only nearly uniquely (Gerhard \etal 1997). In other words,
there is likely a finite region in $\Phi$--space within which for each
$\Phi$ there exists a positive \df\ that matches even exquisite
kinematic data.  It is not practical at present to optimize the
gravitational potential non-parametrically. Thus in work sofar a
functional form with one or two parameters has been assumed, and
confidence limits were obtained for the values of these
parameters. See Section 4 for an example.  Then one must, of course,
assess whether the results are dependent on the chosen
parametrization.

\def\chish{\chi^2_{\sigma+h4}}
\def\chis{\chi^2_{\sigma}}
\def\chih{\chi^2_{h4}}

\section{Results sofar: The E0 galaxy NGC 6703, and others}

In this Section we discuss the results obtained sofar. We will
concentrate on the case of NGC 6703 which we have analyzed (Gerhard
\etal 1997), but mention briefly the results obtained for NGC 2434 by
Rix \etal (1997). Both galaxies are E0, so presumably nearly spherical,
which simplifies the analysis.

The kinematic data for NGC 6703 were obtained during three observing
runs at the Calar Alto 3.5-m telescope with a Boller \& Chivens longslit
twin spectrograph, during a total of 18.5 hrs of observations. A major
axis spectrum and two spectra perpendicular to the major axis and shifted
by 24'' and 36'' from the center were taken. The instrumental resolution
was $85\kms$. The data were reduced as described by Bender (1990) and
Bender, Saglia \& Gerhard (1994); extensive Monte Carlo simulations were
used to test the effects of continuum fitting and instrumental resolution,
and to determine the error bars. The data for $\sigma$ and $h_4$ are
shown in Fig.~\ref{n6703} below.

\begin{figure}[h]
\vspace*{0cm}
\hbox to \hsize{\psfig{file=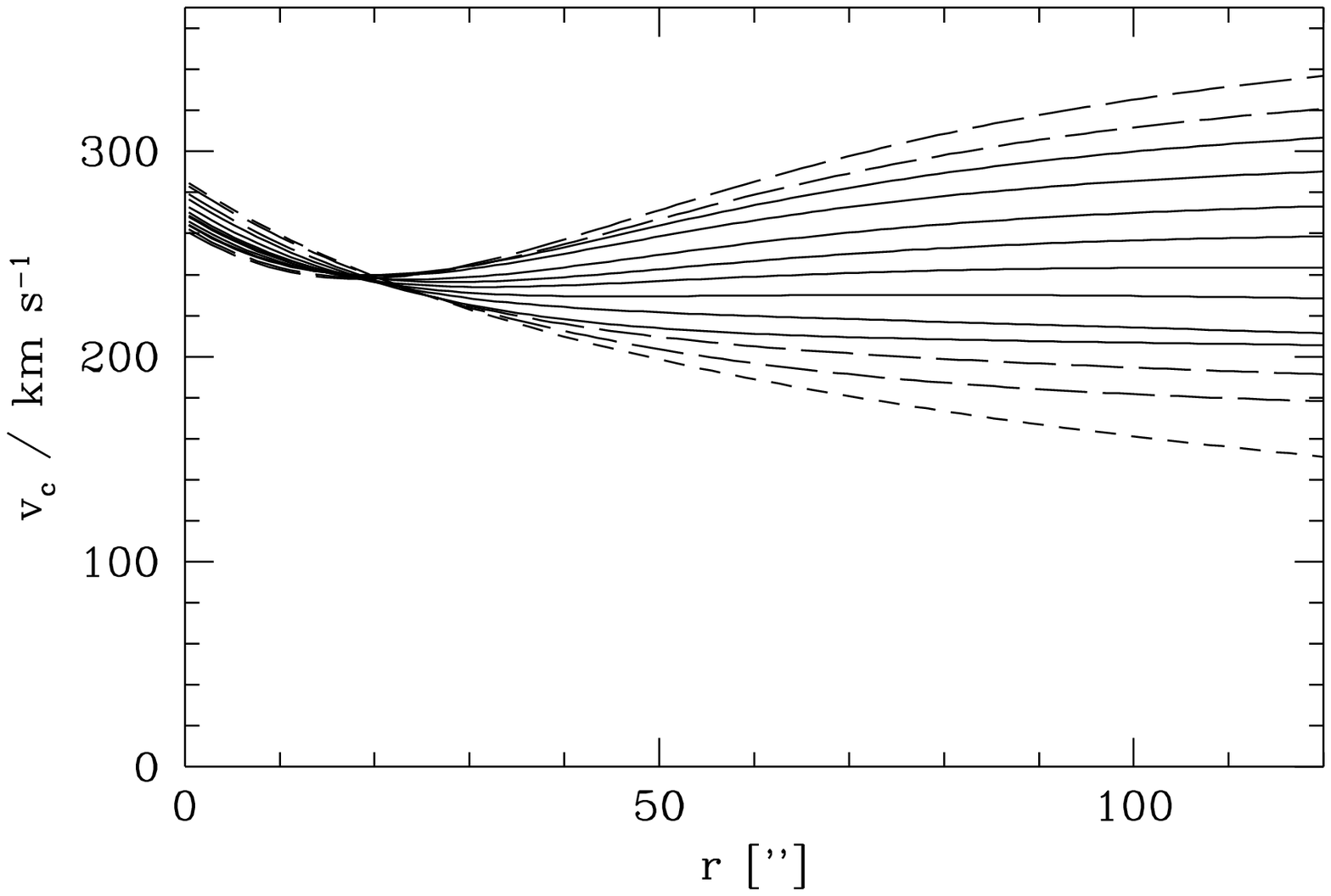,width=\halfwidth} \hfill
		\psfig{file=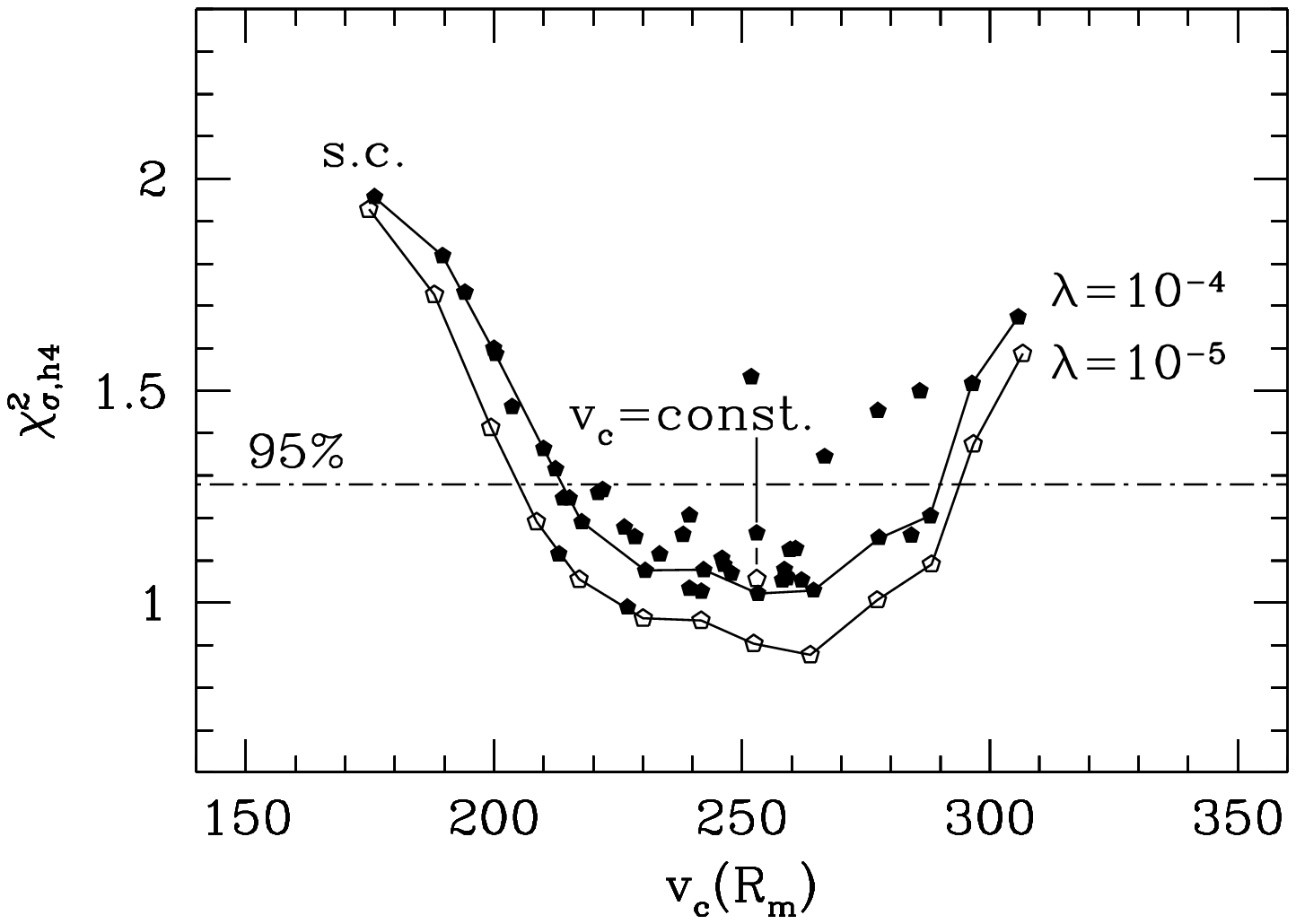,width=\halfwidth} }
 \vspace*{-0.1cm}
 \caption[allvc.eps]{(Left): \ninepoint Rotation curves for a sequence
	of gravitational potentials (stars plus dark halo) used in the
	analysis of NGC 6703.  The full lines show rotation curves
	that are consistent with the NGC 6703 kinematic data inside
	the $95\%$ confidence boundary for $\lambda\equal 10^{-5}$ (open
	symbols in Fig.~\ref{chisqvc}). The other line styles show
	rotation curves inconsistent with the data; among these is the
	constant--M/L model with no dark halo (short--dashed).}
 \label{allvc}
 \vspace*{-0.2cm}
\caption[chisqvc]{(Right): \ninepoint Quality with which the
	kinematics of NGC 6703 can be fitted in different
	potentials. The figure shows the average $\chi^2$ per
	$\sigma$-- and $h_4$ data point, of the best--estimate
	distribution function fitted to the velocity dispersion and
	line profile data. This is plotted as a function of the
	assumed potential's circular rotation velocity at the observed
	radius of the last kinematic data point. Filled symbols show
	best--estimate models derived with the optimal $\lambda\equal
	10^{-4}$; open symbols represent models derived with
	$\lambda\equal 10^{-5}$.  The self--consistent ($M/L\equal
	{\rm const.}$) and the $v_c\equal {\rm const.}$ models are
	marked separately. The horizontal dashed line shows the $95
	\%$ confidence boundary derived from Fig.~\ref{distrchisq}. }
 \label{chisqvc}
\end{figure}

Dynamical models were fit to these data along the lines described in
the previous section. For the luminous component we used a Jaffe model
which fits the surface brightness data to $\lta 0.1$ mag and the
cumulative mass profile to $\lta 2\%$. Models in which this luminous
component has constant M/L and contains all the mass in NGC 6703, fail
to fit the data by a large margin in $\chish$ (Fig.~\ref{chisqvc}).
Either the velocity dispersion profile is fit approximately, but then
the line profiles cannot be reproduced, or the $h_4$ profile is fitted
well, but then $\sigma(R)$ is poorly matched. The non-parametric
algorithm finds a bad compromise which is ruled out at high
statistical significance.

Models with dark matter were investigated using a two-parameter family
of dark halo potentials and rotation curves:
\begin{equation}
\Phi_H(r) = \textstyle{1\over 2} v_0^2 \ln(r^2+r_c^2),
\qquad\quad
v_c(r) = v_0 {r\over \sqrt{r^2 + r_c^2}}.
\end{equation}

\ni These halo mass models have a constant density core, and
the parameters are chosen such that the halo matters only at large
radii. This is similar to the maximum disk hypothesis in spiral
galaxies. We therefore call these luminous plus dark matter models
{\sl maximum stellar mass models}; we show a sequence of such models
in Fig.~\ref{allvc}.

Dynamical models for NGC 6703 were derived from the data in a variety
of these maximum stellar mass potentials, including those shown in
Fig.~\ref{allvc}.  Because the dispersion profile is falling, only one
of the two halo parameters can be determined. We have found that a
well--constrained parameter is the circular orbit velocity at the
radius of the last kinematic data point,
$v_c(R_m)$. Fig.~\ref{chisqvc} shows the results. A self--consistent
potential is clearly inconsistent with the data, while a model with a
completely flat rotation curve is marginally consistent with the data
at $\simeq 1.5\sigma$. All models that are not ruled out at $95\%$
confidence (according to Fig.~\ref{distrchisq}) have $v_c(R_m)=250\pm
40\kms$ at $R_m=2.6 R_e$. This corresponds to a total mass in NGC 6703
inside $78''$ ($13.5\,\hfifty^{-1} \kpc$, where $h_{50}\equiv H_0 / 50
{\rm km/s/Mpc}$) of $1.6-2.6 \times 10^{11} \hfifty^{-1} \msun$, and
to an integrated B-band mass--to--light ratio out to this radius of
$\Upsilon_B=5.3-10$, rising from the central $\Upsilon_B=3.3$ by at
least a factor of $1.6$.

Fig.~\ref{n6703} shows the predicted kinematics from several models
together with the data, and the implied anisotropy profiles.  The
anisotropy of the stellar distribution function in NGC 6703 thus
changes from near-isotropic at the centre to slightly radially
anisotropic ($\beta\equal 0.3-0.4$ at 30'', $\beta\equal 0.2-0.4$ at
60''). It is not well-constrained at the outer edge of the data, where
$\beta\equal -0.5 - +0.4$, depending on variations of the potential in
the allowed range. Notice that large anisotropy gradients between 40''
and 70'' and across the outer data boundary are implied if the
rotation curve of NGC 6703 were completely flat. The model with
$v_c\equal {\rm const}$ is thus not a very plausible one for this
galaxy.

\begin{figure}[t]
\centering
\psfig{file=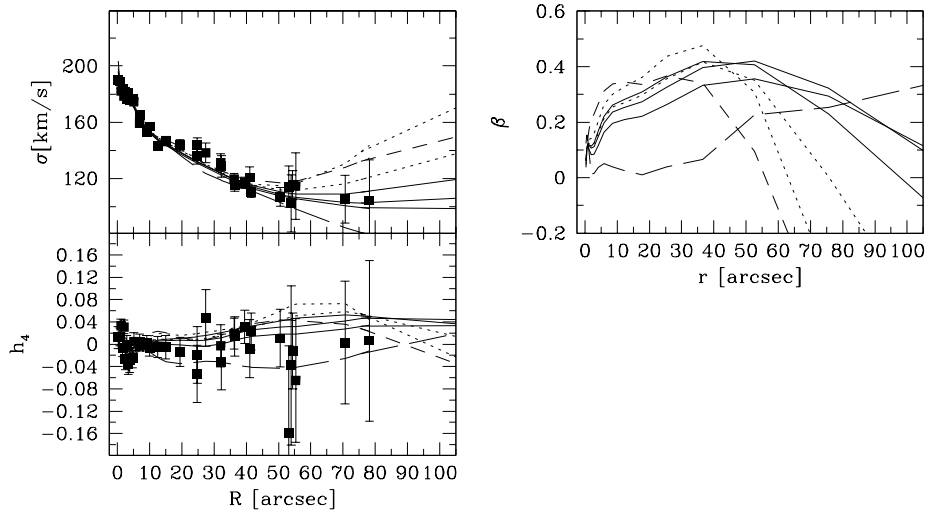,width=\hsize}
 \vspace*{0cm}
 \caption[Dynamical models for the kinematics of NGC 6703] {\ninepoint
    Dynamical models for the kinematics of NGC 6703 in several
    luminous plus dark matter potentials, compared to projected
    velocity dispersion (top left panel) and \vp-shape parameter $h_4$
    (bottom left). The right panel shows the models' intrinsic
    anisotropy parameter $\beta(r)$, with the same linestyles:
    self--consistent model (stars only; long--dashed), $v_c\equal {\rm
    const.}$ model (short--dashed), three models with $v_c(78'')$ in
    the lower part of the acceptable range, according to
    Fig.~\ref{chisqvc} (full), and two models with $v_c(78'')$ in the
    upper part of this range (dotted lines).}
 \label{n6703}
\end{figure}

Rix \etal (1997) have analyzed the velocity profiles of the E0 galaxy
NGC 2434 with a linear orbit superposition method. This galaxy
provides an interesting contrast to NGC 6703 because it has an
essentially flat dispersion profile. Its kinematics are likewise
inconsistent with a constant--M/L potential, but are well--fit by a
model with $v_c\equal {\rm const}$. This can be interpreted as a
maximum stellar mass model in the sense defined above, in which the
luminous component with maximal $\Upsilon$ contributes most of the
mass inside $R_e$. In this case, $\Upsilon\simeq 6.3$.  The kinematics
of NGC 2434 are also well--fit by a range of specific, cosmologically
motivated mass models (Navarro, Frenk \& White 1997) which, if
applicable, would imply lower $\Upsilon$, down to 4.3, 
and significant dark mass inside $R_e$.

It is clear that the dynamical analysis determines only the
distribution of total mass, but cannot fix the mass--to--light ratio
of the stellar component separately. To do this, we must, like in
spiral galaxies, use extra constraints such as from population
analysis.  The fact that the central mass--to--light ratio in NGC 6703
comes out low ($3.3$) even in a maximum stellar mass model may mean
that the contribution of the halo to the mass within $0.5-1R_e$ cannot
be very large. This needs to be studied in more detail; if true it would
give an interesting constraint on galaxy formation theories.

\section{Conclusions and Discussion}

The principal result of the work described here is that both the
anisotropy and the mass distribution of an elliptical galaxy out to
$2-3R_e$ can be derived from velocity dispersion and line profile
shape measurements. The two ellipticals analyzed so far do have dark halos,
the resulting circular rotation velocity curves are nearly flat, and
the mass--to--light ratios of the stellar populations have come out
low. In NGC 6703 there are indications for deviations from a flat
rotation curve in the sense expected if the luminous matter has
segregated to the center; these need to be confirmed with more
accurate observations.  Future studies must show whether there are
differences between the shapes of the true circular velocity curves of
elliptical galaxies.

The strength of the dynamical analysis of elliptical halos is that it
simultaneously yields the stellar-dynamical anisotropy structure, a
further strong constraint on formation theories. Moreover, since the
dynamical analysis works best at high surface brightness, it is
best--suited for constraining small halo core radii. The drawback of
the dynamical analysis is that it can hardly be pushed further than to
about $3 R_e$, and that there are problems in deprojecting
non-spherical systems (Gerhard and Binney 1996).  These can, however,
be eased by concentrating on the flattest elliptical galaxies.  For
the E0 galaxy NGC 6703, current data determine the true circular
velocity at the radius of the last kinematic data point to about $\pm
40 \kms$ at $95\%$ confidence, within the framework of spherical
models.  The largest part of the quoted uncertainty is due to the
finite radial extent of the data.

X-ray observations have the advantage that they can yield the mass
distribution out to very large radii. Only few systems have been
studied sofar, however, some of which are cluster brightest
galaxies. A typical resolution limit is $\sim R_e$.  Gravitational
lensing observations can either give an accurate determination of the
mass within an Einstein radius (in the case of strong lensing), or a
statistical determination of the outer mass profiles of distant
ellipticals (in the case of weak lensing).  Lensing observations only
test projected masses, but they have the advantage that dynamical
equilibrium need not be assumed.

We will therefore learn most about ellipticals by combining the
different techniques, emphasizing their strengths and minimizing the
uncertainties, and covering as large a radial range as possible.  For
example, if the gravitational potential of an elliptical galaxy at
several $R_e$ is known from X--ray data, this will substantially
reduce the error in the dynamically derived outer anisotropy. Stellar
dynamical models derived from absorption line kinematics can be
combined with discrete velocity samples at larger radii, such as from
planetary nebulae searches (Arnaboldi \& Freeman 1997), and stellar
dynamical mass--to--light ratios are required for normalisation in
gravitational lensing studies (Kochanek \& Keeton 1997). 

The fact that also ellipticals have dark halos has been clear
from X-ray and other observations and is not too surprising. However,
the detailed study of their dark matter halos is of particular
interest, since at least cluster ellipticals are now believed to be
very old (see Bender et al.\ 1998 and references therein) 
and must have formed when the mean density of the
Universe was still high. It will also be interesting to see
whether the uniformity of ellipticals in the fundamental plane
extends to their detailed dynamical properties, and whether
this is caused by dynamical processes (Merritt 1997) or reflects
the details of the hierarchical formation process.

\acknowledgments
We acknowledge financial support
by the Schweizerischer Nationalfonds
under grants  21-40464.94 and 20-43218.95, and 
by the Deutsche Forschungsgemeinschaft under SFB 375.

\end{document}